\documentstyle[12pt,aaspp4]{article}
\begin{document}

\title{Screw Instability and Blandford-Znajek Mechanism}

\author{Li-Xin Li}
\affil{Princeton University Observatory, Princeton, NJ 08544--1001, USA}
\affil{E-mail: lxl@astro.princeton.edu}
\affil{(November 5, 1999)}

\begin{abstract}

When magnetic field lines thread a rotating black hole's horizon and connect
with remote astrophysical loads, the rotational energy of the black hole can 
be extracted through the Blandford-Znajek mechanism. Due to the rotation of the 
black hole, the magnetic field lines are twisted and toroidal components are 
generated. So poloidal electric currents are induced and the black hole's 
rotational energy is transported to the astrophysical loads through Poynting flux. 
The Blandford-Znajek mechanism has been considered to be a possible process 
for powering extragalactic jets. 

In this paper we show that due to the screw instability of magnetic field, 
the toroidal components of the magnetic field, and thus the poloidal currents, 
cannot exceed the limits given by the Kruskal-Shafranov criterion. This 
significantly lowers the power of the Blandford-Znajek mechanism when the loads
are far from the black hole. So the Blandford-Znajek 
mechanism can only work efficiently within the neighborhood of the black hole. 
The implications of the results for the scenario of extragalactic jets powered 
by the Blandford-Znajek mechanism are discussed.

\end{abstract}


\section{Introduction}

When a magnetic accretion disk surrounds a black hole, the magnetic
field lines frozen in the disk drift toward the black hole as the
disk plasma is slowly accreted onto the black hole. After the accreted 
plasma particles get into the black hole, the magnetic field lines 
which were frozen to the plasma are released and then thread
the black hole's horizon. So the black hole is magnetized and an
approximately stationary and axisymmetric magnetic field is formed
around it (Thorne, Price, \& Macdonald 1986). The existence of a disk is
necessary for confining the magnetic field lines threading the black hole.
If the black hole is rotating, the magnetic field lines threading it
are twisted and toroidal components of magnetic field
and thus poloidal electric currents are generated. If the other ends
of the magnetic field lines connect with remote non-rotating astrophysical loads,
the rotating black hole exerts a torque on the astrophysical loads and
the rotational energy of the black hole is extracted and transported
to the astrophysical loads via Poynting flux.
This is the so-called Blandford-Znajek mechanism (Blandford \& Znajek 1977). For a
long time the Blandford-Znajek mechanism has been considered to be a
plausible way to power the extragalactic jets (Rees et al 1982; Begelman, 
Blandford, \& Rees 1984; Ferrari 1998).
[Recently some people have argued that the Blandford-Znajek mechanism may be less
efficient since the electromagnetic power from the accretion disk may dominate
the electromagnetic power from the black hole (Ghosh \& Abramowicz 1997; Livio, Ogilvie, 
\& Pringle 1999; Li 1999). Blandford and Znajek have also mentioned this possibility
in their original paper (Blandford \& Znajek 1977). But the
situation may be different for a black hole with a geometrically thick disk
(Armitage \& Natarajan 1999).]

For magnetic field configurations with both poloidal components and
toroidal components, the screw instability plays a very important role
(Kadomtsev 1966; Bateman 1978; Freidberg 1987). If the toroidal magnetic field is 
so strong that from one end to the other end the magnetic field lines wind around 
the symmetry axis once or more, the magnetic field
and the plasma confined by it become unstable against long-wave mode
perturbations, and the plasma column quickly twists into a helical shape.
To maintain a stable magnetic field structure around the
black hole, which is necessary for the working of the Blandford-Znajek
mechanism, the toroidal components of the magnetic field and the poloidal
currents cannot exceed the limits set by the Kruskal-Shafranov
condition which is the criterion for the screw instability (Kadomtsev 1966;
Freidberg 1987).

Recently Gruzinov has shown that if some magnetic field lines connect a black hole
with an accretion disk, the black hole's rotation twists the magnetic field
lines sufficiently to excite the screw instability. This can make the black
hole to produce quasi-periodical flares, and is argued to be a new mechanism
for extracting the rotational energy of black hole (Gruzinov 1999).

In this paper we show that the screw instability of magnetic field play a very
important role in the Blandford-Znajek mechanism. To make the magnetic field and the
plasma confined by it safe against the screw instability, the toroidal components
of the magnetic field and thus the poloidal electric
currents cannot be too big, which significantly lower the power of the
Blandford-Znajek mechanism. The screw instability puts a stringent upper
bound to the power of the Blandford-Znajek mechanism if the distance from
the black hole to the astrophysical loads exceeds some critical value.
The implications of the results for the scenario of extragalactic jets powered by the 
Blandford-Znajek mechanism are discussed. It's argued that jets powered by 
the Blandford-Znajek mechanism cannot be dominated by Poynting flux at large
scales and cannot be collimated by their own toroidal magnetic fields.

\section{Power of the Blandford-Znajek Mechanism}
The power of the Blandford-Znajek mechanism is
\begin{eqnarray}
    P \approx {\Omega_F\left(\Omega_H - \Omega_F\right)\over 4 \pi}
        r_H^2 B_n \Psi_H,
    \label{power}
\end{eqnarray}
where $\Omega_H$ is the angular velocity of the black hole,
$\Omega_F$ is the angular velocity of the magnetic field lines,
$r_H$ is the radius of the black hole horizon, $B_{n}$ is the poloidal
magnetic field on the black hole horizon (on the horizon the poloidal magnetic
field has only normal components), and $\Psi_H \approx B_{n}\pi r_H^2$
is the magnetic flux through the northern hemisphere of the black hole
horizon (Macdonald \& Thorne 1982).
(Throughout the paper we use the geometric units with $G=c=1$.)

On the horizon of the black hole, the toroidal magnetic field $B_H$ is related to 
the poloidal electric current $I$ flowing into the horizon by
\begin{eqnarray}
    B_H \approx {2I\over r_H},
    \label{bt}
\end{eqnarray}
suppose the magnetic field is aligned with the rotation axis of the black hole.
The current is related to the poloidal magnetic field on the horizon by
\begin{eqnarray}
    I \approx {1\over 2}(\Omega_H-\Omega_F) r_H^2 B_n.
    \label{current}
\end{eqnarray}
Inserting Eq.~(\ref{current}) into Eq.~(\ref{bt}), we obtain a
relation between the toroidal magnetic field and the poloidal magnetic
field on the horizon
\begin{eqnarray}
    B_H \approx (\Omega_H - \Omega_F) r_H B_n,
   \label{btp}
\end{eqnarray}
which can also be written as
\begin{eqnarray}
    \Omega_F \approx \Omega_H - {1\over r_H}{B_H\over B_n}.
    \label{wf}
\end{eqnarray}

Inserting Eq.~(\ref{btp}) into Eq.~(\ref{power}) and using $\Psi_H\approx
B_{n}\pi r_H^2 $, we get
\begin{eqnarray}
    P\approx {1\over 4} B_n B_H r_H^3 \Omega_F.
    \label{powera}
\end{eqnarray}

\section{The Screw Instability of Magnetic Field}
For a cylindrical magnetic field with both poloidal and toroidal components (the
so-called screw pinch configuration), the kink safety factor is defined as
\begin{eqnarray}
    q = {{2\pi R B_\parallel}\over L B_\perp},
    \label{safe1}
\end{eqnarray}
where $L$ is the length of the cylinder, $R$ is the radius of the cylinder,
$B_\parallel$ is the poloidal component of the magnetic field which is parallel
to the axis of the cylinder, and $B_\perp$ is the toroidal component of the
magnetic field which is perpendicular to the axis of the cylinder. The screw
instability turns on when
\begin{eqnarray}
    q < 1,
    \label{insta}
\end{eqnarray}
which is called the Kruskal-Shafranov criterion (Kadomtsev 1966, Freidberg 1987).
Though this criterion is usually proved only in flat spacetime, Gruzinov has shown
that it also holds for the force-free magnetosphere around a Kerr black hole
(Gruzinov 1999).

The screw instability is very important for tokamaks and pinches (Bateman 1978).
Since it is a kind of long-wave mode instability, the screw instability can quickly
disrupt the global structure of magnetic field.

Now consider a quasi-cylindrical magnetic field configuration above the northern
hemisphere of the black hole horizon. Then the kink safety factor is given by 
Eq.~(\ref{safe1}). If the magnetic flux and electric current in the cylinder are 
conserved, we have
\begin{eqnarray}
    B_\parallel R^2 \approx B_n r_H^2, \hspace{1 cm}
    B_\perp R \approx B_H r_H.
    \label{conserve}
\end{eqnarray}
Then the kink safety factor can be expressed with $B_n$ and $B_H$
\begin{eqnarray}
    q = {{2\pi r_H B_n}\over L B_H}.
    \label{safe}
\end{eqnarray}
From Eq.~(\ref{safe}) and Eq.~(\ref{insta}), for the magnetic field to be
safe against the screw instability (thus we require $q>1$), we must require
\begin{eqnarray}
    B_H < {2\pi r_H\over L} B_n.
    \label{btu}
\end{eqnarray}
Eq.~(\ref{btu}) gives an upper bound on the toroidal components of the
magnetic field for a stable magnetic configuration. Using Eq.~(\ref{bt}),
the constraint on the induced poloidal electric current is
\begin{eqnarray}
    I < \pi {r_H^2\over L} B_n.
    \label{currentu}
\end{eqnarray}
From Eq.~(\ref{wf}) and Eq.~(\ref{btu}), the
constraint on the angular velocity of the magnetic field lines is
\begin{eqnarray}
    \Omega_H - {2\pi \over L}<\Omega_F<\Omega_H,
    \label{wfu}
\end{eqnarray}
where $\Omega_F < \Omega_H$ is required by that energy is
extracted from the black hole.

The angular velocity of the magnetic field lines (and thus the toroidal
components of the magnetic field and the induced poloidal electric current)
is determined by the rotation of the black hole and the inertia of the remote
astrophysical loads. In the optimal case, i.e. when the power takes its
maximum, $\Omega_F=\Omega_H/2$ (which is called
the impedance matching condition) (Macdonald \& Thorne 1982). From Eq.~(\ref{wfu}),
for the impedance matching case the magnetic field is stable only if 
$L<4\pi/\Omega_H=8\pi r_H(M/a)$, where $M$ is the mass of the black hole and 
$a$ is the angular momentum per unit mass of the black hole [$\Omega_H =
a/(2 M r_H)$]. In other words, the impedance matching condition can be achieved 
only if $ L < 8\pi r_H \left({a\over M}\right)^{-1} $.

\section{Upper Bounds on the Power of the Blandford-Znajek Mechanism}
From Eq.~(\ref{powera}), Eq.~(\ref{btu}), and $\Omega_F<\Omega_H$, we obtain
an upper bound on the power of the Blandford-Znajek mechanism immediately: 
$P<{1\over 4}B_n^2 r_H^3 \Omega_H\left(
{2\pi r_H\over L}\right)$. With more detailed analysis [Using Eq.~(\ref{wf})
for $\Omega_F$ instead of simply replacing $\Omega_F$ with $\Omega_H$ in
Eq.~(\ref{powera})], we obtain
\begin{eqnarray}
    P < {2\over\alpha}\left(1-{1\over 2\alpha}\right) P_0 
    \hspace{1 cm} (\alpha>1),
    \label{poweru2}
\end{eqnarray}
where
\begin{eqnarray}
    \alpha \equiv {L\over 8\pi r_H}{a\over M},
    \label{alp}
\end{eqnarray}
which measures the distance that the Blandford-Znajek process can work on;
\begin{eqnarray}
    P_0 \equiv {1\over 64}\left(a\over M\right)^2 B_n^2 r_H^2,
    \label{pzero}
\end{eqnarray}
which is the power of the Blandford-Znajek process in the optimal case ($\Omega_F = 
\Omega_H/2$). [Eq.~(\ref{pzero}) differs from
the result of Macdonald et al (Thorne, Price, \& Macdonald 1986) by a 
factor of $1/2$ but this is not important for our current purposes.] 

When $\alpha<1$ [i.e. $L<8\pi r_H (M/a)$], the impedance matching condition can be
achieved and thus the upper bound on the power is given by $P_0$
[Eq.~(\ref{pzero})]. Therefore, we obtain the upper bounds on the power of 
the Blandford-Znajek mechanism
\begin{eqnarray}
     P_{\rm max} = \left\{\begin{array}{ll}
                   {2\over\alpha}\left(1-{1\over 2\alpha}\right) P_0
                        \hspace{1 cm} &(\alpha>1)\\
                   P_0 \hspace{1 cm} &(\alpha<1)
                   \end{array}
     \right..
     \label{pmax}
\end{eqnarray}
If $\alpha \gg 1$, $P_{\rm max}\approx {2\over\alpha}P_0$,
the power of Blandford-Znajek mechanism is significantly
lowered by the screw instability of magnetic field (Fig.~\ref{figure1}).

The power limit arising from the screw instability can also be stated as follow:
for a given power $P$ the Blandford-Znajek process can only operate within the
distance 
\begin{eqnarray}
    L<8\pi r_H \left({a\over M}\right)^{-1}f(\eta), \hspace{1 cm}
    f(\eta)\equiv{1\over\eta}\left(1+\sqrt{1-\eta}\right),
    \label{lmax}
\end{eqnarray}
where $\eta\equiv P/P_0$ is the efficiency of the Blandford-Znajek mechanism
($0<\eta\le 1$ always). Eq.~(\ref{lmax}) shows an interesting anti-correlation
between the power of the central engine (black hole) and the maximum distance that 
the central engine can work at. For the Blandford-Znajek 
process to work efficiently (i.e. $\eta\approx 1$), the distance $L$ cannot be 
larger than $ 8\pi r_H \left({a\over M}\right)^{-1}$ (Fig.~\ref{figure1}).

\section{Implications for Extragalactic Jets Powered by the Blandford-Znajek 
Mechanism}
Extragalactic jets are usually thought to be
associated with very powerful energetic processes in active galactic nuclei (AGN). 
The Blandford-Znajek process associated with a rapidly rotating supermassive
black hole has been considered to be a possible mechanism for 
powering extragalactic jets since it provides a practical way for extracting 
the vast rotational energy of the black hole (Rees et al 1982; Begelman, 
Blandford, \& Rees 1984).

From the discussions in previous sections, the screw instability of
magnetic field prevents the Blandford-Znajek mechanism from working at large
distances from the black hole. For the Blandford-Znajek mechanism to work efficiently
(i.e. $\eta = P/P_0 \approx 1$), the distance within which the Blandford-Znajek
mechanism works must satisfy [see Eq.~(\ref{lmax})]
\begin{eqnarray}
    L < 8\pi r_H \left({a\over M}\right)^{-1}
       \approx 2.4\times 10^{-3} {\rm pc} \left({M\over 10^9 M_{\sun}}\right)
       \left({a\over M}\right)^{-1}.
    \label{length}
\end{eqnarray}
Typical extragalactic jets have lengths from several kiloparsecs to several
hundred kiloparsecs (Bridle \& Perley 1984), which are much larger than the limit
given by Eq.~(\ref{length}). 
If the Blandford-Znajek process really works in AGN, it can only work within the
distance limited by Eq.~(\ref{length}) unless it has an extremely low efficiency. 
Beyond that distance, either magnetic flux
or poloidal current (or both) must fail to be conserved within the tube of
fluid lines. [Remember that as we move from Eq.~(\ref{safe1}) to Eq.~(\ref{safe})
we have only used the conservation of magnetic flux and poloidal current.] 
This implies that the magnetic field and poloidal current originating from the
central black hole cannot extend to large distances
in the jets, the astrophysical loads in the Blandford-Znajek mechanism have to
be located at a very short distance from the central black hole [the distance is 
limited by Eq.~(\ref{length})]. So, the bulk of Poynting flux from the central 
black hole is converted into the kinetic energy of 
plasma particles at the very beginning of jets --- in fact the distance given by 
Eq.~(\ref{length}) is too close to the black hole to be resolved with current
observations (Junor, Biretta, \& Livio 1999). The jets so produced cannot be 
dominated by Poynting flux at large radii and cannot be collimated by their own 
magnetic fields.
Therefore, the screw instability of magnetic field gives significant constraints
on the scenario of jets powered by the Blandford-Znajek mechanism.

The transition from a Poynting flux dominated jet to a matter dominated one could 
happen very close to the central black hole, but the collimation of such a jet at 
large radii might be problematic. Though current observations cannot tell if extragalactic
jets are Poynting flux dominated, some observations show that much of the particle 
acceleration occurs at $1-10$~pc from the central black hole (Heinz \& Begelman 1997).

\section{Conclusions}
Considering the screw instability of magnetic field, the Blandford-Znajek
mechanism can only work within a finite distance from the black hole. Beyond 
that distance, either the conservation of poloidal current or the conservation 
of magnetic flux (or both) must break down -- otherwise the screw instability
comes in and the global magnetic field structure is disrupted. The distance 
from the black hole to the boundary within which the Blandford-Znajek mechanism can
work is
\begin{eqnarray}
    L_{\rm max} \approx 8\pi r_H \left({a\over M}\right)^{-1}\left(
    {P\over P_0}\right)^{-1} \left(1+\sqrt{1-{P\over P_0}}\right),
    \label{lmax1}
\end{eqnarray}
where $P\le P_0$ always. If $P\approx P_0$, $L_{\rm max}\approx
8\pi r_H \left({a\over M}\right)^{-1}$. Thus, the Blandford-Znajek mechanism 
can only work efficiently in the close neighborhood of the black hole.

In applying our results to extragalactic jets, we have found that the screw instability 
significantly constrains the scenario of extragalactic jets powered by the 
Blandford-Znajek mechanism. The jets produced by the Blandford-Znajek mechanism
cannot be dominated by Poynting flux at large distances from the central
engine and cannot be collimated by their own magnetic fields.

Though people have discussed various magnetohydrodynamic instabilities for the 
propagation and acceleration of jets (Appl \& Camenzind 1992; Begelman 1998 and 
references therein), nobody has considered the constraint on the power of the central 
engine. The results in this paper have shown that the screw instability of magnetic field
significantly constrains the power of the central engine. The discussions can also be
extended to the case when jets are produced by electromagnetic processes associated
with an accretion disk (Blandford \& Payne 1982). Since the radius of disk is usually 
larger than the radius of black hole, and jets produced by disk is less likely to be 
Poynting flux dominated due to that the coronae of accretion disk are full of mass, 
the jets produced by accretion disk seem to be less vulnerable to the screw instability 
than the jets produced by black hole.

\acknowledgments{I am grateful to Bohdan Paczy\'nski, Paul Wiita, Christian Fendt,
and Julian Krolik for helpful discussions. I am also grateful to the anonymous 
referee for valuable comments. This work was supported by the NSF grant 
AST-9819787.}


\newpage

\figcaption[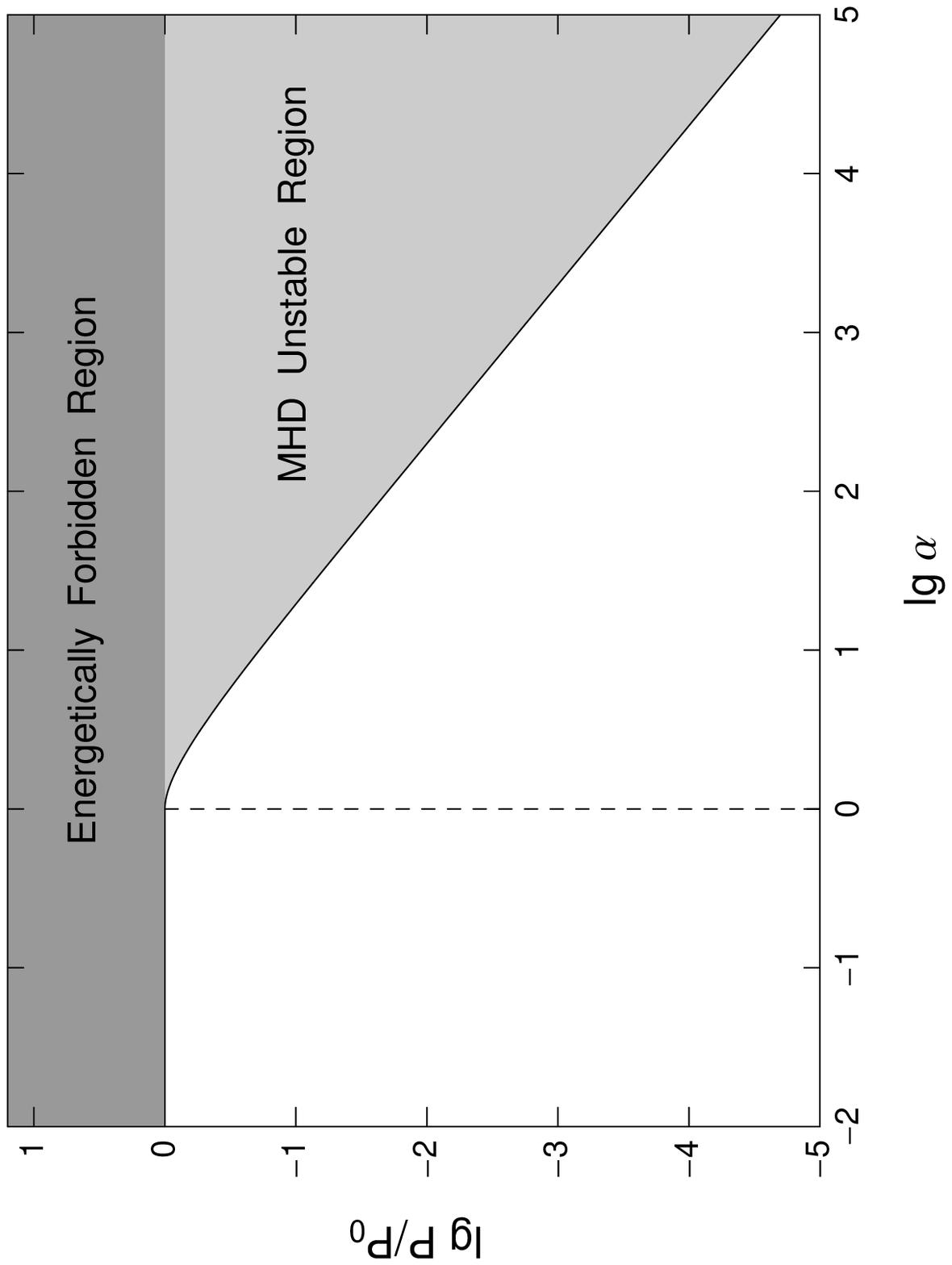]{The restriction on the power $P$ of the Blandford-Znajek
mechanism. The horizontal axis is the logarithm of $\alpha\equiv La/(8\pi
r_HM)$ where $L$ is the distance from the black hole to the astrophysical loads, 
$r_H$ is the radius of the black hole horizon, $M$ is
the mass of the black hole, $a$ is the angular momentum per unit mass of 
the black hole. The vertical axis is the logarithm of $P/P_0$ where $P_0\equiv 
P(\Omega_F=\Omega_H/2)$. The power takes its maximum at $\Omega_F=\Omega_H/2$
(the impedance matching condition), thus $P_0$
gives a universal upper bound on the power of the Blandford-Znajek mechanism.
The region with $P>P_0$ is shown as thick grey region and marked with 
``energetically forbidden region''. For $\alpha>1$, the screw instability gives a
stringent upper bound on the power. The thin grey region marked with ``MHD
unstable region'' is forbidden by the requirement that the magnetic field is
stable against the screw instability. The boundary between the grey regions and the
bright region is the upper bound on $P/P_0$, which is given by
Eq.~(\ref{pmax}) in the text. The Blandford-Znajek mechanism can only work within the bright
region. From the figure we see that, the screw instability significantly restricts
the power of the Blandford-Znajek mechanism when $\alpha>1$. (The boundary
with $\alpha=1$ is shown as a dashed line.) For the Blandford-Znajek mechanism 
to work efficiently (i.e. $P/P_0\approx 1$), we must have $\alpha<1$
[i.e. $L<8\pi r_H (M/a)$].
\label{figure1}}

\end{document}